\documentclass[aps,prd,showpacs,amssymb,amsmath,amsfonts,superscriptaddress,
twocolumn,floatfix,preprintnumbers,altaffilletter]{revtex4}
\usepackage{graphicx}
\usepackage{hyperref}
\usepackage{amssymb}
\usepackage{amsmath}
\usepackage{longtable}
\usepackage{rotating}
\usepackage{color}
\usepackage{epsfig}
\usepackage{epsf}
\usepackage{fancyhdr}

\def\etal{{\it et al.}}

\newcommand{\Lagr}{\mathcal{L}}
\begin{document}
\pagestyle{fancy}
\rhead[]{}
\lhead[]{}
\title{Effect of Correlations Between Model Parameters and Nuisance
Parameters When Model Parameters are Fit to Data
}
\author{Byron Roe}
\affiliation{University of Michigan} 
\date{\today}

%
%
\begin{abstract}
 The effect of correlations between model parameters and nuisance
parameters is discussed, in the context of fitting model parameters
to data.  Modifications to the usual $\chi^2$ method are required.
Fake data studies, as used at present, 
will not be optimum.  Problems will occur 
for applications of the Maltoni-Schwetz~\cite{ms} theorem.
Neutrino oscillations are used as examples, but the problems
discussed here are general ones, which are often not addressed.
\end{abstract}

\preprint{MiniBooNE-304}
\maketitle
%
%
\section{Introduction}
\label{sec:introduction}
Correlation between background parameters and model parameters 
occurs if there are off-diagonal
elements in the overall covariance matrix for background plus
signal or if there is a 
direct dependence between a model parameter and a background. 
For example, in neutrino
oscillation experiments, an initially $\nu_\mu$ beam oscillates producing
some $\nu_e$ events.  The $\nu_\mu$ beam becomes smaller as some of
the $\nu_\mu$ have oscillated.  In some models this disappearance is
considerably larger than the $\nu_e$ appearance.  If there is disappearance
of the $\nu_\mu$, then $\pi^0$ production by $\nu_\mu$, a principle
$\nu_e$ background, also decreases.  The correlation coefficient 
for this example is negative.
Fits often include estimates for this kind of correlation by
using the full covariance matrix and/or modifying the background 
estimates after each iteration.

However, correlation of background and signal can also occur simply
because the
shape of the background spectrum matches the shape of the signal.
They are correlations only in the sense that
the spectrum over bins of the signal is similar to that of the
background and the correlation is found by a regression analysis.  
These correlations should, perhaps, be distinguished by being called 
``regression correlations''.
The present report emphasizes this kind of correlation which is often 
not considered at all.  Modifications to $\chi^2$ fitting are required for
these regression correlations.   If
the correction for these regression correlations is not included, then
the experimental $\chi^2$ will not have a  $\chi^2$ distribution.  The
best fit point then will be incorrect and confidence regions will be too
large.  Even fake data studies will not give correct results if these
corrections are not included and the ``effective
number'' of degrees of freedom obtained from fake data studies will be
affected.
Correction for this new correlation 
makes the experimental  $\chi^2$ distribution be more like the
 $\chi^2$ distribution with number of degrees of freedom equal to the
number of bins minus the number of parameters fit.

\section{Notation}
\label{sec:notation}
Suppose data have been obtained for a histogram with $N_{bins}$.  
The $i$th bin has $N_{data}(i)$ events.

The model used for fitting the data has two kinds of parameters.  
The first kind are nuisance
parameters which are called
``systematic errors''.  There are $N_{bkrd}$ of these parameters.  They
constitute backgrounds, $B_j(i)$ where
$i= 1,...,N_{bins}$, and $j= 1,...,N_{bkrd}$,  The backgrounds have
been evaluated in independent 
experiments previously giving  estimates for the
mean value of each $(B_m)_j(i)$ and the covariance matrix of the error in the
mean value $cov_{B_j}$.  There are correlations from bin to bin, so that the
covariance matrix is not, in general, diagonal.
The total background is
\begin{align}
B^{tot}_m(i) =  \sum_{j=1}^{N_{bkrd}} [(B_m)_j(i)].
\end{align}
Define the ``signal'' as
\begin{align} 
N_{sig}(i) \equiv N_{data}(i) - B^{tot}_m(i).
\end{align}

%
The covariance matrix of the signal in bins $i,j$ is given as 
\begin{align}
cov_{sig}(i,j) = cov_{data}(i,j) + cov_{B^{tot}}.
\end{align}

The second kind of parameters are parameters of the physics model and are
initially unknown.  They are to be fitted to the data, $N_{sig}(i)$.  There
are $N_t$ of these parameters, $t_k,\ k=1,...,N_t$.  The expected value of
the signal in each bin, $N_{th}(i)$, is a function of these parameters.

The fit done is to minimize the $\chi^2$, where 
\begin{align}
\chi^2 = \sum_{i,j} [N_{sig}(i)-N_{th}(i)][cov^{-1}_{sig}(i,j)]
[N_{sig}(j)-N_{th}(j)],
\end{align}
where $[cov^{-1}_{sig}(i,j)]$ is the inverse of the signal covariance matrix.

\section{A toy model}
Suppose one is fitting a parameter to a signal model
of the  form $T(i)=tf(i)$, where $t$ is a constant to be fit
and $f(i)$ is
a known function of the bin number $i$. Suppose, further, that there is
a single background  $B(i)$, which is uncorrelated with the
model parameter.
The fit is done by using the method of Equation 3.  For a $\chi^2$ fit,
the numerator in each bin is 
$[N_{data}(i)-(B_m)(i)-N_{th}(i)]^2 = [N_{data}(i)-(B_m)(i) -tf(i)]^2$.
The denominator is the covariance for the numerator.  The statistical 
covariance of the data
is the uncertainty expected for the current values of $T(i)$ and $B_m(i)$ which
is just $T(i)+B_m(i)$.  In addition there is a term for the uncertainty in
$B_m,\ cov_{B(i)}$.  The denominator  is $T(i)+B_m(i)+cov_{B(i)}$
Let $t_0$ be the
result of that fit. The error is assigned to the $t$ parameter 
by the usual $\Delta \chi^2$ method.

Now suppose that the background has the form $bf(i)$, where $b$ is a constant
and $f(i)$ is the same function that appeared for the model parameter.  The
model and background are now completely correlated.  Since $f(i)$ is a 
common factor for model and background, the numerator for the $\chi^2$ fit
can be written as 
$[N_{data}-(t+b)f(i)]^2= [N_{data}-zf(i)]^2$, where $z\equiv t+b$.  
The background experiment has previously obtained
a mean and error for the parameter $b$.  The fit here obtains a mean and
error for the parameter $z=t+b$.

Assume, incorrectly, that the $\chi^2$ denominator remains the same as for
the uncorrelated fit.  Then,
if $b$ is estimated by $(b_m)$, the new fit for $t$ is the same
as the old one, $t_{fit}=t_0$.

However, the variance matrix is not the same as it was for the previous fit.
For this new fit, the background is part of the
parameter $z$ being fitted. There is no mention of $b$; 
only $z$ appears in the fit. The term $cov_{B(i)}$
is not included in the denominator of the $\chi^2$ term.
In terms of likelihood,
\begin{align*}
\Lagr(data|t,b)\Lagr(b) = \Lagr(data|t+b)\Lagr(b)=\Lagr(data|z)\Lagr(b),\\
\int \Lagr(data|z)\Lagr(b)db =\Lagr(data|z). 
\end{align*}
 The probabilities for $z$ and $b$ are independent. 
When the fit is obtained, the estimate of 
the error in $z$ is done in the usual $\Delta \chi^2$ manner.  The estimate
of $t$ is taken as the difference between $z$ and the
mean value for the background, $(b_m)$. If the fit has gone to the
same place, the result again will be $t_0$.  The error in the estimate
of $t$ is obtained in the usual way by adding the errors
in $z$ and in $(b_m)$ in quadrature. This means that the uncertainty in $t$
will also be the same as it would be if the correlation were ignored.

Since the $\chi^2$ in this fit does not have $cov_{B(i)}$ as part
of the denominator of the $\chi^2$ fit, then, if the fit goes to the
same place, the $\chi^2$ is larger than the $\chi^2$ would have been if
the correlation were not taken into account
and the probability of the fit is lower.  
The probability of the fit is systematically overestimated if the correlation
is ignored. The subtlety here is that this is a fit.  If a fixed value
of $t$ is used and there is no fit, the $\chi^2$ distribution would be correct 
if the  $cov_{B(i)}$ is included.
However, the fitted value of $t$ is not affected by the  $cov_{B(i)}$, and
the $\chi^2$ for the fitted value is too low.  

Note that, in practice, some uncertainties affect both the signal and
background in the same way.  For the MiniBooNE experiment, 
errors in scintillation fraction
and errors in pion production cross sections are of this form.  These
errors should be included in the covariance used in the fit.

For the MiniBoone experiment, does the $\pi^0$ background resemble the signal? 
$\pi^0$
background numbers for the neutrino exposure are obtained
from Table 6.5 of the MiniBooNE Technical Note 194~\cite{oneninetyfour}.
The neutrino spectrum and the neutrino quasi-elastic cross sections were
read from Figures 2 and 5 from a MiniBooNE neutrino elastic scattering
Physical Review article~\cite{mbelastic}.  
Results are shown in Figure 1.  For
$\Delta m^2= 2$ or $1$ eV$^2$, the mean and $\sigma$ of the  $\pi^0$ background
and the signal are quite close and the correlations are large. 

Another method to treat this problem would be to make a combined fit for
the background and signal using the histograms for both signal and
background.  However, if there are many backgrounds, the ``curse of
dimensionality'' makes this impractical.  In addition the many backgrounds
may have been obtained by many different methods.  
 
In the next two sections, the problem of partial regression
correlations of signal
with one or more backgrounds at a given model point will be treated. 
Complications due to dependence on model
parameters of these quantities and other problems
will be discussed in sections VI-VIII.


\begin{figure}
\includegraphics[height=11.1cm]{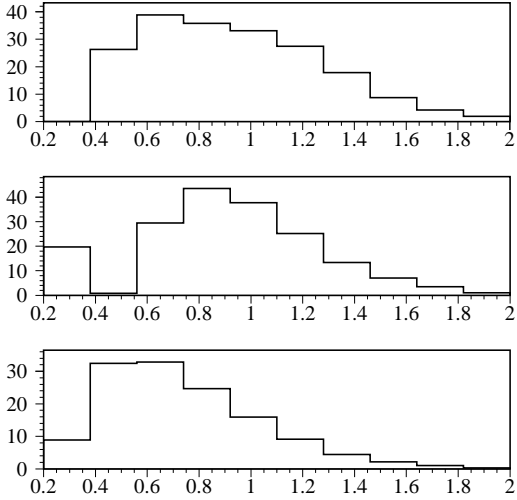}
\caption{The top figure shows the energy spectrum expected
for the  $\pi^0$ background.  The mean is 0.940 GeV and the standard
deviation is 0.341 GeV.  The middle figure shows the energy spectrum
expected for the neutrino oscillation signal if $\Delta m^2 = 2$ eV$^2$.
The mean is 0.910 GeV, the standard deviation is 0.343 GeV and
the correlation with the $\pi^0$ spectrum is 0.718.  
The bottom figure shows the energy spectrum
expected for the neutrino oscillation signal if $\Delta m^2 = 1$ eV$^2$.
The mean is 0.739 GeV, the standard deviation is 0.305 GeV and the
correlation with the $\pi^0$ spectrum is 0.771.
} 
\label{Figure 1}
\end{figure}
\section{Partial correlations}
\label{sec:partial correlations}
In practice, regression correlations are almost never complete; 
it is necessary to 
consider partial correlations between signal and background.
The model for the theory is taken as $tf(i)$, where $t$ is a constant and
$f(i)$ is a known function of bin number.  
Assume that the only significant regression correlation is the one
between the model and $B_k(i)$. 
The model for the background is taken
as $B_k(i)=b_kg(i)$, where $b_k$ is constant and $g(i)$ is a known function of 
bin number.
Let $s$ denote the predicted signal given $t$. 
Let $M_f$ and $\sigma_f$ be the mean and standard deviation 
of $f(i)$ and  let $M_g$ and $\sigma_g$  be the mean and 
standard deviation of the background, $g(i)$. The variation here is
the variation over the bins of the histogram.
Let $x_s^*(i)$ be the part of the 
background correlated to the signal.  Consider a plot of background
versus theory, where the points are the values for each bin. A straight
line regression
fit of background on signal yields
\begin{align}
\frac{x_s^*(i) - M_g}{\sigma_g} = \rho\frac{f(i) - M_f}{\sigma_f}.
\end{align}
The regression correlation coefficient is given by
\begin{align}
\rho = \frac{E\{(g - M_g)(f-M_f)\}}{\sigma_g\sigma_f},
\end{align}
where $E$ means the expectation value over bins.

Since this was a straight line fit $x_s^*$ has the same bin dependence
as the signal. Note that Equations 5 and 6 are
 normalization independent because of
the divisions by $\sigma$.
\begin{align} 
x_s^*(i) = M_g + \frac{\rho\sigma_g}{\sigma_f}\left(f(i) - M_f\right).
\end{align}
Define an ``effective'' correlation 
$\rho_{eff} = \rho\sigma_g/\sigma_f$. 
Let $\Delta_{sk}(i)$ be the uncorrelated part of the background
\begin{align}
\Delta_{sk}(i) = b_k[g(i)-x_s^*(i)]\nonumber \\ 
=b_k[g(i)-M_g -\rho_{eff}(f(i)-M_f)].
\end{align} 
The covariance matrix due to the uncertainty in the experimental values
obtained for background from the experiments determining the background is
given by:
\begin{align}
cov_{\Delta_{sk}}(i,j) = [g_k(i)-M_g-\rho_{eff}(f(i)-M_f)] \nonumber \\
[g_k(j)-M_g-\rho_{eff}(f(j)-M_f)]\sigma^2_{b_k}.
\end{align}
The new covariance matrix for the fit 
uses only this uncorrelated part of $B_k$: 
\begin{align}
cov_{new} = cov_{data} + cov_{(\Delta_{sk}+\sum_{j\ne k}B_J)}.
\end{align}
The fit is then made to
\begin{align} 
N_{fit}(i)  
= (t+b_k\rho_{eff})f(i). 
\end{align}
$N_{sig}$ had subtracted the estimates of all background.
$N_{fit}$ adds to $N_{sig}$ the estimate of the correlated part of $B_k$.
The fitting parameter is taken as
\begin{align}
zf(i)=(t +\rho_{eff}b_k)f(i).
\end{align}
After the fit the estimate for $t$ is found from
\begin{align}
t_{fit} = z_{fit} -\rho_{eff}b_k.
\end{align}
The covariance of $t$ is given by
\begin{align}
cov_t = cov_{z_{fit}} + cov_{\rho_{eff}b_k}.
\end{align}

\section{Correlations of several background parameters with the signal}
It may be that several nuisance parameters have significant 
regression correlation with the signal.
One needs to
find the regression correlation of the signal with the totality of the nuisance
parameters.  This correlation can be calculated by 
standard methods~\cite{probbk}.\\
Let
$\Lambda\equiv (\lambda_{ik})$ be the moment matrix over bins
for the nuisance parameters
with  $t$ added as an extra parameter.
$\Lambda^{ik} =$ the $i,k$ cofactor of $\Lambda$, that is 
$ (-1)^{i+k}\times $ the determinant of $\Lambda$ with
row $i$ and column $k$ removed.\\
Define the correlation matrix:
\begin{align}
 P \equiv(\rho_{ik}) =\left( \lambda_{ik}/[\sigma_i\sigma_k]\right).
\end{align}
Note that, as was the case for Equation 6, $P$ is independent of the
normalization.
For a variable $y$ use $y'=y -M_y$. 
Let 
\begin{align}
(B')_s^*(i) \equiv\sum_{k\ne s} \beta_{sk}B'_k(i).  
\end{align}
where $\beta_{sk}$ are constants chosen to
minimize the expectation value over the bins
\begin{align*}
\frac{1}{N_{bins}}\sum_i[(B_s^*)'(i)-(S')(i)]^2\},
\end{align*} 
where $S(i)$ 
is the predicted number of
data minus background events, given $t$.  It can be shown that
\begin{align}
\beta_{sk} = -\Lambda^{sk}/\Lambda^{ss}=-\sigma_sP^{sk}/(\sigma_kP^{ss}).
\end{align}
$\beta_{sk}$ is not independent of the normalization of the
backgrounds. The method finds the appropriate {\it linear} sum of backgrounds
which has the highest  correlation with the signal. Problems
can arise if non-linear effects are important, as will be noted in 
section VIII.   
Go back from primed coordinates to unprimed coordinates.  The mean of
$B_s^*$ is $M_{B^*_s} =\sum_{k=1}^{N_{bkrd}}\beta_{sk}M_{B_k}$, where $ M_{B_k}$ is the
mean over the histogram bins of the backgrounds $B_k$.  $B^*_s(i)$ can
now be taken as an effective single background for determining the
correlation with signal.  $B_s^*(i)$ will not, in general, have the same
dependence on bins as the signal. As was the case for the one background
case, a linear regression correlation $x_s^*(i)$ 
between the signal and the $B^*_s$ for fixed values of the model
parameters is needed. Suppose the model depends on one parameter and the
dependence is 
given as before by
$tf(i)$ where $f(i)$ is known.  
The regression line of  $B^*_s$ on signal yields
\begin{align} 
\rho^* =\frac{\lambda_{f{B_s^*}}}{\sigma_f\sigma_{B_s^*}},
\end{align}
\begin{align} 
\lambda_{f{B_s^*}} =\frac{1}{N_{bins}}\sum_{i=1}^{N_{bins}} \nonumber \\
\left( \sum_{k=1}^{N_{bkrd}}\beta_{sk}[B_k(i)-M_k][f(i)-M_f]\right).
\end{align}
\begin{align}
   \frac{x^*_s(i) -M_{B_s^*}}{\sigma_{B_s^*}} = \rho^*\frac{f(i)-M_f}{\sigma_f}
\end{align}
or
\begin{align} 
x^*_s(i) = M_{B_s^*} +D(i)\lambda_{f{B_s^*}};
\ \ D(i)=\frac{f(i)-M_f}{\sigma^2_f}.
\end{align}
The uncorrelated part of the background is
\begin{align}
\Delta^*(i) = \sum_{k=1}^{N_{bkrd}} B_k(i) - x^*_s(i) = \nonumber \\
 \sum_{k=1}^{N_{bkrd}} B_k(i)- M_{B^*_s} -D(i)\lambda_{f{B^*_s}}
\end{align}
\begin{align}
\Delta^*(i) = \sum_{k=1}^{N_{bkrd}} B_k(i)\nonumber \\
 -\frac{1}{N_{bins}}
\sum_{k,j}\beta_{sk}B_k(j)\left(1+D(i)[f(j)-M_f]\right) \nonumber \\
+\frac{1}{N^2_{bins}}\sum_{k,j,m} \beta_{sk}B_k(m)D(i)[f(j)-M_f].
\end{align}
Next the covariance of $\Delta^*(i)$ with respect to uncertainties in the 
$B_k(i)$ as determined in preliminary background experiments is calculated.
\begin{align}
cov_{\Delta^*(i)}= T1 +T2 + T3, 
\end{align}
\begin{align}
T1 = \sum_{k,\ell}cov[B_k(i)B_\ell(i)] \nonumber \\ 
-\frac{2}{N_{bins}}\sum_{k,\ell,j}\beta_{s\ell}
(1+D(i)[f(j)-M_f])cov[B_k(i)B_\ell(j)] \nonumber \\
+\frac{2}{N_{bins}^2}\sum_{k,\ell,j,m}D(i)(f(j)-M_f)\beta_{s\ell}
cov[B_k(i)B_\ell(m)],
\end{align}
\begin{align}
T2 = \frac{1}{N_{bins}^2}\sum_{k,\ell,j,m}\beta_{sk}\beta_{s\ell}
[1+D(i)]^2[f(j)-M_f] \nonumber \\ 
[f(m)-M_f]cov[B_k(j)B_\ell(m)] \nonumber \\
-\frac{2}{N_{bins}^3}\sum_{k,\ell,j,m,n}\beta_{sk}\beta_{s\ell}
[1+D(i)]D(i)  \nonumber \\
[f(j)-M_k][f(m)-M_f]cov[B_k(j)B_\ell(n)],
\end{align}
\begin{align}
T3 = \frac{1}{N_{bins}^4}\sum_{k,\ell,j,m,n,p}\beta_{sk}(j)\beta_{s\ell}(m)
[D(i)^2 \nonumber  \\
[f(j)-M_f][f(m)-M_f]cov[B_k(n)B_\ell(p)].
\end{align}
\begin{align}
cov_{new} = cov_{data} + cov_{\Delta^*_{sB^*}}
\end{align}
Since, by construction, $x_s^*(i)$ has the bin dependence of $f(i)$,
define $b^*$ by 
\begin{align}
x_s^*(i) = b^*\rho^*_{eff}f(i), \ {\rm where}\ 
\rho^*_{eff} = \rho^*\frac{\sigma_{B^*_s}}{\sigma_f}
\end{align}
The calculation then proceeds in analogy with that given in the previous
section, Equations 11-14, with the obvious substitutions.

\section{Multiple Model Parameter Calculations}
\label{sec: Multiple Model Parameter Calculations}
For multiple model parameters, it is necessary to apportion the
correlated part of the background among the fitting parameters $z_k$.  For one
parameter we had:\\
\begin{align*}
zf(i)=(t +\rho^*_{eff}b^*)f(i).
\end{align*}
Here $f(i)$ is the spectral shape of the model parameter as a function of bins
in the histogram.  Now there are several model parameters $t_kf_k(i)$ and
corresponding fit parameters $z_k$.

Note that the apportionment is only necessary after the fit since
each parameter $z_kf_k(i)$ will have the spectral shape $f_k(i)$ of the
model parameter.  The overall
$\chi^2$ can be obtained without knowing the apportionment.  This has
the important advantage that the apportionment procedure need be done only
once, not at each stage of the fit.

If the model spectrum is a simple sum of the terms for the 
various parameters, a 
method similar to that used for several
backgrounds is followed. Let
$\Lambda_{model}\equiv ([\lambda_{model}]_{ik})$ be the moment matrix for 
the model parameters determined from the spectrum over the histogram bins
with  the background distribution $(x_{b}^*)$ added as an extra parameter.
Here the previous $x^*_s$ has been retitled $x^*_b$ for clarity in the 
present section. 

Let
$[\Lambda_{model}]^{ik} =$ the $i,k$ cofactor of $\Lambda_{model}$, that is 
$ (-1)^{i+k}\times $ the determinant of $\Lambda_{model}$ with
row $i$ and column $k$ removed.\\
Define the correlation matrix: 
\begin{align} 
P_{model} \equiv \left((\rho_{model})_{ik}\right) =
\left( [\lambda_{model}]_{ik}/\left([\sigma_{model}]_i[\sigma_{model}]_k
\right)\right).
\end{align} 

Let 
\begin{align}
\alpha_{bk} = -\Lambda_{model}^{bk}/\Lambda_{model}^{bb} \nonumber \\
=-[\sigma_{model}]_bP_{model}^{bk}/([\sigma_{model}]_kP_{model}^{bb}).
\end{align}

The $\alpha_{bk}$ are the apportionment parameters.
These parameters may need an overall renormalization so that 
\begin{align}
(x')_b^*(i) =\sum_{k\ne b} \alpha_{bk}t_kf'_k(i),  
\end{align}
where the primed variables have been defined to have zero mean, 
$y'= y -M_y$; $M_y$ is the mean of $y$. 
\begin{align}
z_k(i) =(1+\alpha_{bk})t_kf_k(i).
\end{align}  

However, often the model parameters are not just simple sums.
Consider the
simple neutrino oscillation fit, used in MiniBooNE. 
The two parameters $\Delta m^2$ and $\sin^22\theta$ occur as a
product, not a sum.  The
latter parameter is just a scale parameter determining the size of the
effect.  All of the spectral shape information is contained in $\Delta m^2$.
In terms of the notation $tf(i)$, $t$ is determined by  $\sin^22\theta$
and $f(i)$ by $\Delta m^2$.  The two parameters work together to produce
a single spectrum.  For a background  which matches the shape for a given
 $\Delta m^2$, any mis-estimate of the background will appear in
the fitted value of $\sin^22\theta$ and the entire correlated part of the
background should be associated with that parameter.  
\begin{align}
z_{\sin^22\theta} f_{\sin^22\theta}(i)= (1+\alpha_{b\sin^22\theta})
t_{\sin^22\theta}f_{\sin^22\theta}(i)
\end{align}

The individual terms in the sum of model terms will sometimes be these
composite terms.  This occurs, for example in fits of neutrino data 
for sterile neutrino 
hypotheses. There will sometimes also be more complicated dependences than the
simple ones here and it is necessary to examine the situation for each
particular experiment.  These same considerations apply to the backgrounds
treated in the previous section. 

\section{Some further complications}
\label{sec: Some further complications}
For simplicity use the conditions for Section IV, one background parameter,
one model parameter for this discussion.  Generalization to more
general situations is straight forward.

Suppose that there is a systematic uncertainty which is correlated between
background and data.  The fit for $z$ includes the data uncertainty
(perhaps a normalization uncertainty), but is otherwise unchanged.  However, 
when one is finding the uncertainty in $t=z-\rho_{eff}b$, 
there is a correlation between the uncertainty in $z$ and 
in $b$ which must be taken into account.

The background is $B(i)=bg(i)$.  Until now it has been assumed that
the spectrum function $g(i)$ is fixed.  Suppose there is an uncertainty
in $g(i)$.  This causes an uncertainty  in $\rho$ which
can be calculated using standard error propagation techniques. 
 The
uncertainty in $\rho$ will introduce an uncertainty in the fraction
of background subtracted from data for the fit to $z$.  This
uncertainty must be included
in the covariances used to fit $z$.  This does change the $\chi^2$ of the
fit for $z$. The uncertainty in $\rho$ must also be included in the
uncertainty for $t=z-\rho_{eff}b$.

\section{Incorporating these correlations in practice}
\label{Incorporating these correlations in practice}
In practical situations it 
often occurs that
the backgrounds and correlations vary with the 
model parameters.   An
appropriate fitting procedure for $\chi^2$  fits needs to include these
point-to-point changes in correlations.

For the best fit, the regression correlation should be evaluated at
each step just as the effects of changes in systematic errors are often
evaluated at each step.  The  $\chi^2$ for the best fit is then the lowest
 $\chi^2$ obtained and has the regression correlation correction
appropriate to that best fit set of parameters.

For determining confidence regions, consider a parameter point A. If the
absolute $\chi^2$ is to be used and A
is a fixed point, not the result of a fit, then the regression correlations
should not be included in the calculation of $\chi^2$.  However, if
the $\Delta \chi^2$ method is used, this is a comparison of
 two values obtained in a fit. 
 The $\chi^2$ at A using the regression
correlations at A minus the best fit  $\chi^2$ using the
regression correlations at the best fit should then be used.

If a fake data study is used, the procedure is similar.  There is no
change in the choice to the usual procedure for choosing Monte Carlo events.  
The regression
correlations do not enter and only the usual backgrounds are randomly varied.

What happens next depends upon the question asked.  Suppose there is no fit
and the question
asked is, ``What is the distribution  of $\chi^2$ if the model parameters
are fixed?''  For this question, the regression correlations do not enter.

However, if there is a fit, then 
for each fake data MC example, the calculations are done as
indicated above for point A and for the best fit using the 
regression correlations.  Then, the procedure
for the fake data study is done as before, just counting the fraction
of MC samples having a higher  $\chi^2$.

If the model and background correlations vary with the model point and
the regression correlations are ignored, the best fit point will be
different.  Furthermore, confidence regions will be too large even for the
fake data studies, since the
width of the experimental $\chi^2$ distribution 
will include the variances from the full backgrounds, rather than
just those from the uncorrelated parts of the background.  
At present, the use of an "effective number of degrees of
freedom" is frequently employed to give corrections for the direct
calculations, but not for the fake data studies. That correction is
far less precise than the procedure studied here. 
With this new procedure the
effective number should be closer to the real number (although it may
still useful to include the new effective number as a residual
correction).


\section{The Maltoni-Schwetz theorem}
\label{sec:Maltoni-Schwetz theorem}
Suppose one is examining the compatibility
of two sets of data for a given hypothesis, data set 1 and data set 2. 
Suppose one finds 
the best fits for the model parameters for data set 1, data set 2,
and for the combined data sets 1 plus 2. Let the number of parameters
fit for the three fits be $N_1,\ N_2$ and $N_{1+2}$. The
Maltoni-Schwetz theorem~\cite{ms} then states that
\begin{align} 
\chi^2_{MS}\equiv \chi^2_{1+2} - \chi^2_{1} - \chi^2_{2}.
\end{align}
is a measure of the compatibility of the data sets. If they are
compatible, $\chi^2_{MS}$ has a $\chi^2$ distribution 
with $N_{MS}=N_{1}+N_{2}-N_{1+2}$ degrees of freedom. 

Consider, as an example,
the question of explaining some possible neutrino experiment anomalies as
being due to the presence of two sterile neutrinos.  The model for this
hypothesis assumes several $\sin^22\theta$-like variables, several
$\Delta m^2$-like variables and a $CP$ phase. The first data 
set corresponds to the
appearance of  $\nu_e$ events from an originally $\nu_\mu$ beam 
and the second data set is for the
disappearance of $\nu_\mu$ events from an originally $\nu_\mu$ beam.  

Fits have been made using both $\nu_e$ appearance and $\nu_\mu$ disappearance
experiments~\cite{jc}~\cite{jk}.  Both kinds of experiments are fit
reasonably well with this model, but, using the Maltoni-Schwetz formalism,
tension is found between the appearance and the disappearance experiments. 

For the J. M. Conrad \etal\ fits~\cite{jc},
the number of fitted variables for each of the three data sets was
$N_{app} = 5;\ \ N_{dis} = 6;\ \ {\rm and}\ \ N_{comb} = 7$.  This leads to
$N_{MS} = 5+6-7=4$.
For the disappearance experiments there is no $\pi^0$ background.
There is a  $\pi^0$ background for some of the appearance experiments
including the MiniBooNE experiment which has a large weight within the
appearance experiment sample.
In previous sections it was found
that, if the correlations with backgrounds were not taken into account, 
the $\chi^2$ for the appearance experiment was lower than it should be.

In addition, because of the sensitivity of the appearance experiment to the
$\pi^0$ background, an error in the estimate of that background
can have a disproportionate effect~\cite{jctwo}. For the
combined data fit, if the $\pi^0$ background is larger than the value 
estimated, some $\sin^22\theta$ type parameters will want to be bigger 
than they should be
for the appearance bins, but not for the disappearance bins, giving some
tension within the combined data set, i.e., increasing the $\chi^2$.  
Furthermore the number of degrees of freedom for MS is only four, which
makes the discrepancy turn into an extremely low probability.  The
MS method is especially sensitive to these errors. 
Even without the
correlations considered here, much
of the tension between the appearance and disappearance results goes away
if it is assumed that
the MiniBoone estimate of the $\pi^0$ background is low by 1.4 $\sigma$.

\section{Summary}
\label{sec:summary}
Methods are given for using the $\chi^2$ method when correlations between
nuisance parameters and parameters being fitted occur.
These methods  
are appropriate whenever these correlations exist.
\begin{enumerate}
\item{} If nuisance parameter-signal shape correlations are not taken into
account, $\chi^2$ fits will systematically overestimate the probability of the 
fit. The experimental $\chi^2$ will not have a $\chi^2$ distribution.
 
\item{} Fake data studies without including these correlations will not
be optimum.  The use of ``effective number of degrees of freedom'' 
will help the
situation, but will not be as precise as the methodology introduced here.

\item{} One must use caution in applying the Maltoni-Schwetz theorem to find the
compatibility of two sets of data to a model hypothesis.  The theorem 
may indicate incompatibility if there are correlations
between the nuisance parameters and the signal and/or if there
are problems with the estimations of nuisance variables.
\end{enumerate} 

\section{Acknowledgements}
\label{sec:acknowledgement}
I wish to acknowledge the considerable help of Zuming-He, the H. C. Carver
Professor of Statistics, University of Michigan, during several
discussions which helped clarify
the problem and correct several errors.  MIT graduate student 
Christina Ignarra and
LANL Staff Member Zarko Pavlovic went through the note in detail in order
to apply it to MiniBooNE, and in the process, helped
to find ambiguities and errors. I also wish to thank MIT Professor 
J. M. Conrad for pointing
out to me the effect of changing $\pi^0$ background
on the Maltoni-Schwetz probabilities for a two sterile neutrino picture.    


{}

\end{document}